\documentclass[prb,twocolumn]{revtex4-1} 

\usepackage{amsmath}  
\usepackage{amsfonts} 
\usepackage{graphicx} 

\usepackage{orcidlink}

\usepackage[group-digits = false]{siunitx}
\usepackage{verbatim}
\usepackage{physics}
\usepackage[version=4]{mhchem}

\usepackage{amssymb}
\usepackage{mathtools}
\usepackage{bbold}

\usepackage{booktabs}
\usepackage{tabularx}
\usepackage{float}

\usepackage[caption=false]{subfig}

\usepackage{tikz}
\usetikzlibrary{patterns}
\usetikzlibrary{arrows.meta}

\usepackage{ifthen}

\usepackage{hyperref}
\hypersetup{colorlinks=true, citecolor=blue, urlcolor=blue, linkcolor=blue}
\usepackage{cleveref}

\newcommand{\e}{\mathrm{e}}

\begin{document}


\title{Learning shape resonances from the stabilization method}


\author{Daniel Kromm\,\orcidlink{0009-0009-2922-6314}}
\email{daniel.kromm@tu-darmstadt.de}
\affiliation{Technische Universität Darmstadt, Department of Physics, Institut f\"ur Kernphysik, 64289 Darmstadt, Germany}
\author{Hans-Werner Hammer\,\orcidlink{0000-0002-2318-0644}}
\email{Hans-Werner.Hammer@physik.tu-darmstadt.de}
\affiliation{Technische Universität Darmstadt, Department of Physics, Institut f\"ur Kernphysik, 64289 Darmstadt, Germany}
\affiliation{ExtreMe Matter Institute EMMI and Helmholtz Forschungsakademie Hessen für FAIR (HFHF),
GSI Helmholtzzentrum f{\"u}r Schwerionenforschung GmbH, 64291 Darmstadt, Germany}
\author{Artem G. Volosniev\,\orcidlink{0000-0003-0393-5525}}
\email{artem@phys.au.dk}
\affiliation{Department of Physics and Astronomy, Aarhus University, Aarhus,
DK-8000, Denmark}
\affiliation{ExtreMe Matter Institute EMMI,
GSI Helmholtzzentrum f{\"u}r Schwerionenforschung GmbH, 64291 Darmstadt, Germany}


\date{\today}

\begin{abstract}
Resonances in quantum mechanics are commonly introduced as quasi-bound states embedded in the continuum, a perspective that can be conceptually challenging due to the abstract nature of continuum states. In this work, we discuss an alternative approach that avoids an explicit treatment of the continuum by formulating the problem in terms of discrete quantum states. Our discussion is based on the stabilization method, in which the system is confined to a finite region such that the continuum is replaced by a discrete energy spectrum. Resonances then appear as characteristic features in the energy levels under variation of the confining box size, providing an intuitive interpretation in terms of a two-level system while remaining closely connected to standard quantum mechanics curriculum. We review the method, derive selected results, and discuss practical strategies for extracting resonance parameters from stabilization diagrams. In addition to established fitting procedures, we introduce a novel approach based on the analysis of spatial localization of resonant states, which enables a robust identification of resonance properties. The approach is illustrated using both attractive and repulsive delta-shell potentials, which serve as simple and instructive model systems amenable to analytical treatment. 
\end{abstract}

\maketitle

\section{Introduction}
Quantum mechanics is essential for engaging with modern quantum technologies, motivating ongoing efforts to improve how it is taught~\cite{Johnston1998,McKagan2008,Pollock2023,Buzzell2025}. One concept that students typically find difficult to grasp is that of a scattering resonance. In standard presentations, this concept is associated with a quasi-bound state that decays via quantum tunneling into the continuum of available states~\cite{Davydov1976,Sakurai2020}. Students often struggle to connect this picture to their prior knowledge, in part because such resonances have only limited analogies in classical mechanics, rendering the underlying concepts less intuitive. A related complication is the need to introduce a continuum of states, which can further obscure the physical interpretation for students encountering these ideas for the first time. These challenges motivate the development of novel approaches to teaching scattering resonances and the underlying physics~\cite{Stec1990,delaMadrid2002,Garg2014,Edison2025,Neulinger2026}.

In this paper, we discuss a framework for introducing scattering resonances using the stabilization method, which avoids explicit treatment of the continuum. 
The pedagogical goal of this approach is to discuss resonances in continuum systems at the beginning of a wavefunction-based (`position-first') quantum mechanics course. 
In particular, the method allows students to (i) identify resonances through simple features of discrete spectra and (ii) relate these features to familiar two-level dynamics. More advanced application of the method allows one to extract resonance parameters using computational tools accessible at the undergraduate level. As such, the framework can serve both as an introduction to resonances and as a bridge to more advanced treatments.
 By framing the resonance problem in terms of discrete quantum states, the proposed approach provides a direct conceptual bridge to the resonances in two-level systems~\cite{Griffiths2018,CohenTannoudji2019} that students typically encounter earlier in `spin-first' quantum mechanics education. The two-level framework may therefore better appeal to the already developed physical intuition of the students. 
 
 We base the discussion on the stabilization method~\cite{hazi_taylor1970}, which is not commonly used in standard physics curricula, even though it is well known in the research community.
For a recent pedagogical discussion of resonances and related theoretical methods, including stabilization methods, see Ref.~\cite{koenig2026resonance}.
Within this framework, we present three complementary approaches for extracting resonance parameters: a direct fit to the finite-volume energy levels, an analysis based on the density of states, and a method based on calculating probabilities. To the best of our knowledge, the latter has not been previously discussed in this context.
We show that this method can be particularly useful when traditional stabilization diagnostics fail.

 The simplest way to incorporate the results of this work into an instructional setting is to assign an exercise in which students construct the stabilization diagram discussed in Sec.~\ref{sec:Delta_shell}.
 This modest addition to a standard curriculum can help motivate the concept of resonance and establish a connection to two-level dynamics. This approach requires only familiarity with ordinary differential equations subject to closed boundary conditions, a topic that is typically covered at the second-year undergraduate level.
 A more advanced option is to use this paper as the basis for small numerical undergraduate projects (with a typical workload of at least 50 hours), in which the methods developed here are applied to potentials known to support resonances. At the graduate level, the present work can serve as an intuitive introduction to resonances and the stabilization method.  

{\color{black}Before proceeding, we note that the pedagogical claims presented here are based on one-on-one interactions with students and have not yet been evaluated in large classroom settings. We also note that the proposed framework is currently being implemented through an undergraduate project at Aarhus University, with the present paper serving as the primary background and reference material for the project.}

This paper is organized as follows: In Sec.~\ref{sec:resonances} we briefly review the basics of the concept of resonance, introduce the stabilization method and illustrate the basic manifestation of resonances in this framework. In Sec.~\ref{sec:continuum}, we explain how to extract properties of resonances (position and width) using the stabilization method. To this end, we connect the problem in a box to a problem in the continuum. We illustrate the method and provide benchmark values in Sec.~\ref{sec:Delta_shell} for both attractive and repulsive delta-shell potentials. Finally we summarize and provide four appendices with technical details for instructors.

\section{Resonances and the Stabilization Method}
\label{sec:resonances}

\subsection{Resonances}

To place scattering resonances in context for students, we briefly recall how the notion of resonance appears in different areas of physics.
Quantum resonances occur when a quantum system exhibits an enhanced response at specific energies or frequencies. During undergraduate studies, students typically encounter this concept in a number of related contexts, which we summarize below in broad terms.

In quantum optics, a quantum resonance occurs when a two-level system (such as a spin-1/2 particle) is coupled to an external field~\cite{Griffiths2018}. This phenomenon shares conceptual parallels with ``resonance'' or ``mesomerism'' in chemistry, where the coupling between degenerate electronic configurations significantly alters the molecular structure and properties~\cite{CohenTannoudji2019}.

In atomic and molecular physics~\cite{Sakurai2020,messiah_quantum_2020}, a resonance occurs when an $n$-level system (e.g., a harmonic oscillator) is coupled to light. This type of resonance can be intuitively visualized using the example of a molecule driven by a single-frequency external electric field. Mathematically, the corresponding physics closely mirrors the classical resonance familiar to students from their studies in classical mechanics~\cite{Taylor2004} and electrodynamics~\cite{Griffiths2023}.

\begin{figure}[t]
    \centering
    \begin{tikzpicture}[scale=1]

    \draw[red, semithick, dashed] (0,0.8) -- (6.6,0.8);
    \node[red] at (5.1,1.2) {quasi-bound state};
    
    \draw[blue, semithick, dashed] (0,-0.8) -- (6.6,-0.8);
    \node[blue] at (5.1,-1.2) {bound state};
    
    \draw[->] (0,0) -- (6.6,0) node[right] {$x$};
    \draw[->] (0,-1.5) -- (0,2.7) node[above] {$V(x)$};
    
    \begin{scope}
        \clip (0,-2.2) rectangle (6.6,3.6);
        \draw[thick, domain=0.26:6.2, smooth, samples=300]
            plot (\x,{0.2/(\x*\x) - 2*exp(-1.4*(\x-1.2)^2) + 1.05*exp(-2.2*(\x-3.0)^2)});
    \end{scope}
    
    \draw[red, thick, domain=0.01:2.7, smooth, samples=50]
        plot (\x,{0.8 - 1 * exp(-0.7/\x) * sin(deg(2.2*\x))});
    \draw[red, thick, domain=2.69:3.31, smooth, samples=50]
        plot (\x,{0.8 + 0.744*exp(-8*(\x-2.337)^2)});
    \draw[red, thick, domain=3.3:6.2, smooth, samples=50]
        plot (\x,{0.8 - 0.05*sin(deg(8*(\x-3.3)))});
   
    \draw[blue, thick, domain=0.01:6.2, smooth, samples=200]
        plot (\x,{-0.8 + 7*exp(-1/\x) * exp(-2*(\x-1.5)^2) * exp(-1.2*\x))});

    \node at (3.5,2.5) {barrier};
    \draw[-] (3.5,2.3) -- (3.1,1.1);

\end{tikzpicture}
    \caption{A schematic potential $V(x)$ supporting both a bound state and a quasi-bound state. The quasi-bound state is temporarily localized in the well region below the barrier, while the finite barrier allows for tunneling, giving rise to resonant behavior.}
    \label{fig:res_potential}
\end{figure}
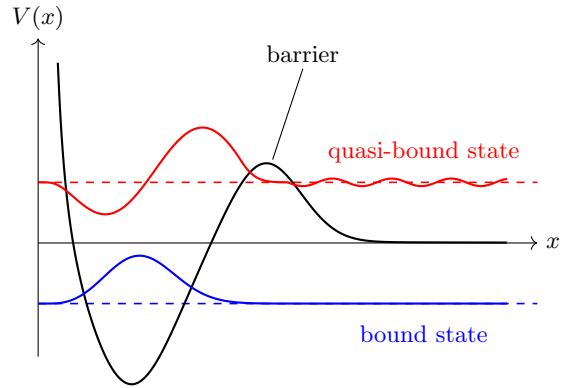

Teaching the concept of resonances in scattering, however, presents a greater challenge. These resonances emerge when a discrete state couples to a continuum, and their characteristics depend on the nature of both the discrete state and the continuum. Different types of resonances -- such as Feshbach or Fano resonances~\cite{Chin:2010crf}  -- can arise in this context. Fortunately, many properties of resonances follow solely from unitarity rather than from the details of the underlying microscopic dynamics~\cite{Weinberg1995}. Therefore, to lay the foundation, textbooks often focus on arguably the simplest family of resonances -- quantum shape resonances~\cite{Sakurai2020} --  which were first introduced by Gamow to explain alpha decay. These are metastable states in which a particle or quasiparticle is temporarily trapped by the shape of a potential barrier (see Fig.~\ref{fig:res_potential}). Unlike bound states, which are permanently localized, shape resonances eventually decay into the continuum. This decay manifests as distinct features -- such as sharp peaks or asymmetric profiles -- in observables like scattering cross-sections, absorption spectra, or transmission probabilities.

\subsection{Shape resonances}

The simplicity of Fig.~\ref{fig:res_potential} (provided the concept of quantum tunneling is understood) makes shape resonances the simplest toy model for introducing quantum resonances in scattering. This model provides students with a foundation to develop their physical understanding of resonance processes.

 Mathematical frameworks for studying shape resonances typically involve concepts from scattering theory, e.g., the S-matrix or phase shifts, where a rapid change in phase (by $\pi$) signals a resonance, {\color{black} see also Sec.~\ref{sec:phase_shift}}. The Breit-Wigner formula is often introduced to model the total scattering cross section, $\sigma$, in the vicinity of the resonance energy\cite{Sakurai2020}:
\begin{equation}
\sigma(E)\propto \frac{1}{E}\frac{\Gamma^2}{(E-E_\mathrm{r})^2+(\Gamma/2)^2}\,,
\end{equation}
where $E$ is the energy, $\Gamma$ is the width of the resonance, and $E_\mathrm{r}$ is the position of the resonance. Advanced textbooks associate poles of the S-matrix with shape resonances~\cite{Taylor:1972pty}. 

Introducing shape resonances through scattering theory requires concepts that often extend beyond the scope of introductory quantum mechanics, which is why most courses only briefly mention them.
Scattering theory itself demands a strong grasp of solving the Schr{\"o}dinger equation for continuum states, presenting both mathematical and conceptual challenges. In particular, scattering theory can be challenging for students, as it involves reasoning about non-normalizable states.

Beyond mathematical challenges, the use of scattering theory can also make the term ``resonance'' itself seem misleading. While students are typically familiar with resonances from classical mechanics -- where the phenomenon arises from the interaction between an intrinsic frequency and a driving frequency -- they often struggle to connect this intuitive picture to the quantum regime. In quantum scattering, resonances involve a continuum of states rather than discrete frequencies, which can make the concept feel abstract and unfamiliar.

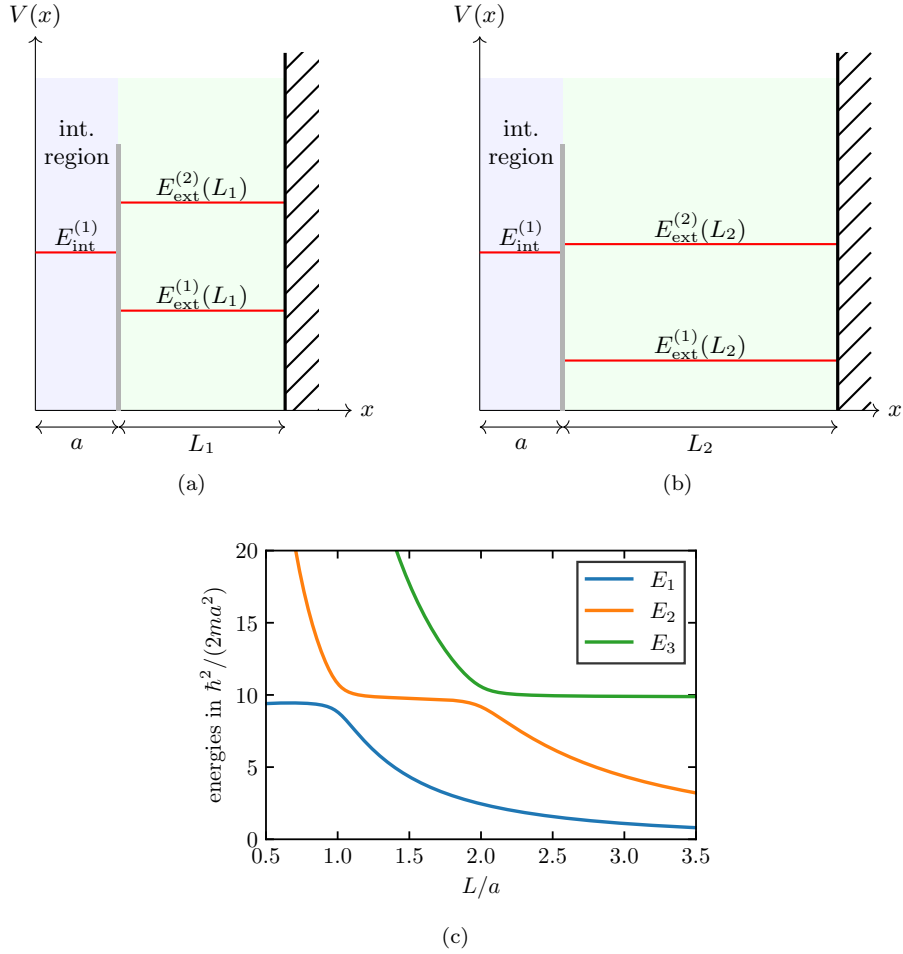
\begin{figure*}[t]
    \centering

    \subfloat[]{%
        \begin{tikzpicture}[scale=1.1]
            \fill[blue!5] (0,0) rectangle (1,4);
            \node[align=center] at (0.5,3.2) {int.\\ region};

            \fill[green!5] (1,0) rectangle (3,4);
            \draw[<->] (1.0,-0.2) -- (3.0,-0.2);
            \node at (2.0,-0.4) {$L_1$};
            \draw[<->] (0.0,-0.2) -- (1.0,-0.2);
            \node at (0.5,-0.4) {$a$};

            \draw[->] (0,0) -- (3.8,0) node[right] {$x$};
            \draw[->] (0,0) -- (0,4.5) node[above] {$V(x)$};

            \draw[red, thick] (0,1.9) -- (1,1.9);
            \node at (0.5,2.1) {$E_\mathrm{int}^{(1)}$};

            \draw[red, thick] (1,1.2) -- (3,1.2);
            \node at (2,1.4) {$E_\mathrm{ext}^{(1)}(L_1)$};

            \draw[red, thick] (1,2.5) -- (3,2.5);
            \node at (2,2.7) {$E_\mathrm{ext}^{(2)}(L_1)$};

            \fill[gray!60] (0.97,0) rectangle (1.03,3.2);

            \draw[very thick] (3,0) -- (3,4.3);

            \begin{scope}
                \clip (3,0) rectangle (3.4,4.3);
                \foreach \y in {-1,-0.7,...,5}
                    \draw[thick] (2.7,\y) -- (3.5,\y+0.8);
            \end{scope}
        \end{tikzpicture}%
    }
    \hspace{0.04\textwidth}
    \subfloat[]{%
        \begin{tikzpicture}[scale=1.1]
            \fill[blue!5] (0,0) rectangle (1,4);
            \node[align=center] at (0.5,3.2) {int.\\ region};

            \fill[green!5] (1,0) rectangle (4.3,4);
            \draw[<->] (1.0,-0.2) -- (4.3,-0.2);
            \node at (2.65,-0.4) {$L_2$};
            \draw[<->] (0.0,-0.2) -- (1.0,-0.2);
            \node at (0.5,-0.4) {$a$};

            \draw[->] (0,0) -- (4.8,0) node[right] {$x$};
            \draw[->] (0,0) -- (0,4.5) node[above] {$V(x)$};

            \draw[red, thick] (0,1.9) -- (1,1.9);
            \node at (0.5,2.1) {$E_\mathrm{int}^{(1)}$};

            \draw[red, thick] (1,0.6) -- (4.3,0.6);
            \node at (2.65,0.8) {$E_\mathrm{ext}^{(1)}(L_2)$};

            \draw[red, thick] (1,2.0) -- (4.3,2.0);
            \node at (2.65,2.2) {$E_\mathrm{ext}^{(2)}(L_2)$};

            \fill[gray!60] (0.97,0) rectangle (1.03,3.2);

            \draw[very thick] (4.3,0) -- (4.3,4.3);

            \begin{scope}
                \clip (4.3,0) rectangle (4.7,4.3);
                \foreach \y in {-1,-0.7,...,5}
                    \draw[thick] (3.9,\y) -- (4.7,\y+0.8);
            \end{scope}
        \end{tikzpicture}%
    }
    \hspace{0.04\textwidth}
    \subfloat[]{%
        \includegraphics[width=0.4\textwidth]{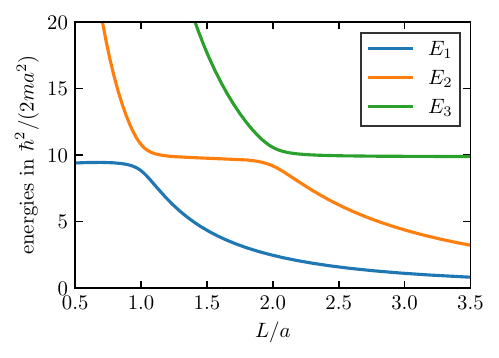}%
    }

    \caption{Illustration of the stabilization method with an interior (interaction) and exterior region. The interior and exterior regions provide energy levels that can couple through the barrier. If one shifts the boundary of (a) to a different position (b), the energy levels are also shifted. In the case of (b) the second level in the exterior region is close to the energy level in the interior region, which allows them to mix. (c) Eigenvalues of the three-level Hamiltonian in Eq.~\eqref{eq:hamiltonian_toy} as a function of $L$. }
    \label{fig:stab_scheme}
\end{figure*}

\subsection{The Stabilization Method}
\label{stab_method_intro}
To address the challenges discussed earlier, we propose introducing resonances using the stabilization method by Hazi and Taylor~\cite{hazi_taylor1970}. This approach allows students to calculate scattering properties within a confined system
enabling the exploration of resonances immediately after studying the particle-in-a-well problem, which is typically introduced early in the course.
For simplicity, we focus here on one-dimensional geometries. However, our choice of the boundary condition (`infinite wall' at the origin) also makes it straightforward to extend the discussion to radial three-dimensional quantum problems, offering a clear path to more advanced applications.

To introduce the stabilization method, we consider 
the stationary Schr\"odinger equation,
\begin{equation}
\left( -\frac{\hbar^2}{2m} \dv[2]{x} + V(x)\right)\psi(x) = E \psi(x)\,,
\label{eq:Schr_equation_1}
\end{equation}
for a particle with mass $m$ in a one-dimensional box problem with two regions, see Fig.~\ref{fig:stab_scheme}. These regions are called interior and exterior region. The interior region contains information
about the physical system, whereas the exterior region contains information about the boundary condition of our problem. The interface between exterior and interior regions can be chosen rather arbitrarily as long as the interior region contains all relevant information about the interaction. 
Of course, one should check that the results are independent of this choice (see supplementary material for an explicit example).

By assumption, the exterior and interior regions are almost disconnected from each other. We imagine a high potential barrier that prevents the coupling of two regions, see Fig.~\ref{fig:res_potential}. To formally introduce the regions, one can start with an infinitely high potential barrier separating them, in which case the two regions are disconnected. In this case, the interior region has states with the energies $\{E_\mathrm{int}^{(1)}, E_\mathrm{int}^{(2)}, ...\}$ whereas the exterior region has the energies
$\{E_\mathrm{ext}^{(1)}(L), E_\mathrm{ext}^{(2)}(L), ...\}$, see Fig.~\ref{fig:stab_scheme}(a),(b). The argument $L$ implies that these energies depend on the width of the exterior region.

When we lower the barrier height, the states from the interior and exterior regions begin to mix. However, as long as the barrier is sufficiently high, this mixing significantly affects the states only when their energies are close. To solve the problem, we consider the three-level problem illustrated in Fig.~\ref{fig:stab_scheme}(a): 
\begin{equation}
H=\begin{pmatrix}
    E_\mathrm{int}^{(1)} & \Delta_1 & \Delta_2 \\
    \Delta_1 & E^{(1)}_\mathrm{ext}(L) & 0\\
    \Delta_2 & 0 & E^{(2)}_\mathrm{ext}(L)
\end{pmatrix}\,,
\label{eq:hamiltonian_toy}
\end{equation}
where the parameters $\Delta_1, \Delta_2\ll E_\mathrm{int}^{(1)}, E^{(1)}_\mathrm{ext}(L),E^{(2)}_\mathrm{ext}(L) $ describe the couplings between the states. {\color{black} We note that the exterior regions do not mix directly, as reflected by the vanishing matrix elements connecting them. This result follows directly from the assumption that the exterior states are described by free plane waves propagating outside the potential region. Consequently, any interaction between the states in the exterior region must be mediated by the interior region.}

The coupling $\Delta_i$ can be calculated using perturbation theory. 
For example, for a delta shell potential
($V(x)=U\delta(x)$, see Sec.~\ref{sec:Delta_shell}), we can use perturbation theory in the limit of {\color{black} an almost
impenetrable barrier} ($U\to\infty$) to derive (see Appendix~\ref{app:strong} {\color{black} for details of the derivation})
\begin{equation}
\Delta_i\simeq -\frac{\hbar^4}{4 m^2 U}\left(\frac{\mathrm{d}\psi^{(1)}_{\mathrm{int}}}{\mathrm{d}x}\frac{\mathrm{d}\psi^{(i)}_{\mathrm{ext}}}{\mathrm{d}x}\right)_{x=a},\quad i=1,2\,.
\label{eq:coupling_strong}
\end{equation}
{\color{black} Note that $\Delta_i$ is proportional to $1/U$, which is small by assumption. Therefore, the coupling between the exterior and interior regions is important only if $|E_{\mathrm{int}}^{(1)}- E_{\mathrm{ext}}^{(i)}|\sim |\Delta_i|$, otherwise it can be neglected. To illustrate this, we diagonalize the Hamiltonian from Eq.~(\ref{eq:hamiltonian_toy}), see Fig.~\ref{fig:stab_scheme}(c).  
Outside of the regions where $|E_{\mathrm{int}}^{(1)}- E_{\mathrm{ext}}^{(i)}|\sim |\Delta_i|$, the figure features} (i) a stabilized state, i.e., a state whose energy ($\simeq E_{\mathrm{int}}^{(1)}$)  does not depend on the parameter $L$; (ii) states of the exterior region whose energies change as $\propto 1/L^2$. 

The stabilized state can be interpreted as a quasi‑bound state, while the exterior states form an effective continuum. For certain values of $L$, the energies of these states become nearly degenerate ($E_{\mathrm{int}}^{(1)}\simeq E_{\mathrm{ext}}^{(i)}$), and their coupling gives rise to avoided crossings.
These avoided crossings represent the hallmark of a resonance and can be naturally connected to the familiar two‑level dynamics of quantum mechanics~\cite{Griffiths2018}. In particular, when the energy separation satisfies
$|E_{\mathrm{int}}^{(1)}-E_{\mathrm{ext}}^{(i)}|\gg\Delta_i$
the system can be prepared exclusively in the exterior state. By contrast, when this condition is not met and the exterior state is populated, one expects Rabi‑like oscillations of the probability density between the interior and exterior regions.

\section{Connection to the problem in continuum}
\label{sec:continuum}

From the discussion above, it is evident that the system’s response in the near‑degenerate case $E_{\mathrm{int}}^{(1)}\simeq E_{\mathrm{ext}}^{(i)}$ differs qualitatively from that in the far‑detuned regime, where $|E_{\mathrm{int}}^{(1)}-E_{\mathrm{ext}}^{(i)}|\gg\Delta_i$. This distinction provides the most elementary definition of a resonance.
To help students connect the finite-box picture to experimentally measurable quantities, it is necessary to connect the solutions of Eq.~(\ref{eq:hamiltonian_toy}) to an appropriate scattering formulation of the problem. To this end, we note that the wave function for a given energy in a box also solves the corresponding continuum problem within the same spatial region. Furthermore, since the solutions to the Schr{\"o}dinger equation in the exterior region are plane waves, we can extend these solutions seamlessly to infinite space. Thus, even though the problem is initially solved within a box, we effectively obtain also the scattering solution at that energy. In other words, for a given $L$ the stabilization method contains full information about scattering at a given energy. By changing the value of $L$, we obtain a direct link between the box-confined and {\it all} scattering solutions. Below, we investigate this link using phase shifts, density of states and quasi-bound state probability. We stress however that while these continuum concepts are used for validation and interpretation, the stabilization method itself requires only bound‑state techniques. Therefore, from an instructional perspective, the stabilization method provides a flexible entry point to scattering resonances that can be adapted to different levels of abstraction.

\subsection{Phase shift}\label{sec:phase_shift}
{\color{black}
To make the connection to the standard scattering theory description of resonances more explicit, we briefly review the concept of the scattering \emph{phase shift} for one-dimensional problems where the wave function vanishes at the origin~\cite{Lifshits1981-kq}.
For Galilei-invariant interactions, the scattering of two particles with mass $m_1$ and $m_2$ reduces to the scattering of a single degree of freedom with the reduced mass $m$ off the interaction potential. If this potential is short-ranged, i.e., if the interaction goes to zero sufficiently fast outside some characteristic interaction range, the degree of freedom has the energy $E=\hbar^2 k^2/(2m)$ and is described by a free wave outside of this characteristic range. However, the presence of the potential modifies the phase of the scattering wave function compared to a free solution of the Schr\"odinger equation. This modification is quantified by the phase shift $\eta(E)$. More precisely, far outside the interior region, where the potential vanishes, the stationary wave function takes the universal form
\begin{equation}
    \psi_k(x)\xrightarrow[x\to\infty]{}\sin\big(kx+\eta(E)\big)\,,
    \label{eq:boundary_condition}
\end{equation}
where $x$ is the relative distance between the particles.
In a typical scattering experiment, one detects the final state particles far away from the interaction region, i.e., for large $x$, where this universal form of the wave function applies.
Thus the phase shift $\eta(E)$ encodes all the information available about the scattering process. In particular, a rapid variation of $\eta(E)$ as a function of energy signals the presence of a resonance because the scattering wave is temporarily trapped in the potential for a time $\Delta t = \hbar\, {\mathrm{d}\eta}/{\mathrm{d}E}$ \cite{Taylor:1972pty}, which follows from the density of states, see below.
Physically, this corresponds to a quasi-bound state that temporarily traps the particle. 
This temporary trapping adds an extra half-cycle (or \(\pi \) radians) to the spatial wave function compared to non-resonant energies.
In more general settings with multiple scattering channels (e.g., even and odd parity), a separate phase shift enters for each channel. However, in the present one-dimensional geometry, only a single phase shift appears.

Near an isolated resonance, the phase shift can be decomposed into a slowly varying background contribution $\eta_\mathrm{bg}(E)$ and a rapidly varying resonant part that changes by $\pi$ across the resonance with the characteristic Breit-Wigner form~\cite{Taylor:1972pty},
\begin{equation}
    \eta(E)=\eta_\mathrm{bg}(E)+\arctan\!\left(\frac{\Gamma/2}{E-E_\mathrm{r}}\right)\,.
    \label{eq:phase_shift}
\end{equation}
The background phase shift $\eta_\mathrm{bg}(E)$ describes the smooth, non-resonant scattering from the potential and typically varies only weakly with energy in the vicinity of the resonance. The second term captures the rapid variation associated with the quasi-bound state. Here, $E_\mathrm{r}$ denotes the resonance energy and $\Gamma$ its width, which is inversely related to the lifetime of the quasi-bound state.}

Within the stabilization method, we do not directly solve the scattering problem in infinite space but instead impose a finite box of size $L$. This leads to a discrete set of eigenenergies $E_N(L)$, where $N=1,2,3,\dots$ labels the eigenstates in ascending order of energy. The discrete eigenenergies are constrained by the scattering phase shift. {\color{black}Indeed, applying the boundary condition $\psi(L)=0$ in Eq.~(\ref{eq:boundary_condition}) leads to the quantization condition}
\begin{equation}
    {\color{black} kL+\eta(E)= n\pi\,,\quad n\in\mathbb{Z}\,.}
\end{equation}
This relation provides a direct link between the finite-volume spectrum and the continuum scattering properties. In particular, changes in the phase shift are reflected in the $L$-dependence of the discrete energies. Focusing on a specific eigenstate $N$, one obtains a curve $E_N(L)$ as the box size is varied, see Fig.~\ref{fig:stab_scheme}. Near a resonance, these curves exhibit plateaus, i.e., regions where the energy changes only weakly with $L$. This behavior can be understood as a consequence of the rapid variation of the phase shift as a function of energy in the resonance region. Assuming that the smooth contribution $n\pi-k(E)L-\eta_\mathrm{bg}(E)$ varies approximately linearly with $L$ in the vicinity of the plateau, one can derive a fitting function for the $N$-th eigenvalue,
\begin{align}
    E_N(L)
    &= E_\mathrm{r}
    + \frac{\Gamma}{2\tan\!\left[
        n\pi-k(E)L-\eta_\mathrm{bg}(E)
    \right]} \notag\\
    &\simeq E_\mathrm{r}
    + \frac{\Gamma}{2\tan\!\big(\tfrac{L-L_N}{\Delta L_N}\big)}\,.
    \label{eq:phase_shift_fit_linearization}
\end{align}
{\color{black} The first line follows directly from Eq.~(\ref{eq:phase_shift}).}
In the second line, the smooth contribution $n\pi-k(E)L-\eta_\mathrm{bg}(E)$ has been approximated linearly in $L$ near the plateau center. Thus, $L_N$ and $\Delta L_N$ are effective fit parameters describing the position and width of the plateau for the $N$-th eigenstate. In practice, one identifies a plateau in $E_N(L)$ and performs a fit of this form around its center to extract the resonance energy $E_\mathrm{r}$ and width $\Gamma$. However, this procedure can be delicate, since it involves a four-parameter fit in a region where the energy varies only weakly. Consequently, it requires high-quality data and is most reliable for narrow resonances, where the plateau is well pronounced, see an illustration in Sec.~\ref{sec:Delta_shell}.

\subsection{Density of States}
An alternative and often more robust way to extract resonance parameters is based on the \emph{density of states} (DOS)~\cite{dos_mandelshtam1993}. Conceptually, the DOS $\rho(E)$ measures how densely energy levels are distributed around a given energy $E$. In other words, $\rho(E)\,\mathrm{d}E$ gives the number of states in an energy interval $\mathrm{d}E$.


In scattering theory, the density of states carries direct physical information about resonances. 
In particular, it can be shown that the DOS is related to the energy derivative of the phase shift. 
This relation can be understood heuristically from the quantization condition 
$kL+\eta(E)=n\pi$, which determines the allowed energies in a finite volume. 
Differentiating with respect to energy, one finds that the spacing between neighboring levels is governed by 
$\partial_E(kL+\eta(E))$, such that the density of states is proportional to
\begin{equation}
    \rho(E)\propto \frac{\partial}{\partial E}\big(kL+\eta(E)\big)
    \simeq \frac{\partial \eta(E)}{\partial E}\,,
\end{equation}
where the smooth contribution from $k(E)L$ has been separated from the rapidly varying phase shift.
As a consequence, a resonance, characterized by a rapid variation of the phase shift, appears as a pronounced peak in the DOS.
For an isolated resonance, this peak has approximately a Lorentzian (Breit-Wigner) shape centered at the resonance energy $E_\mathrm{r}$ with width $\Gamma$ (see Eq.~(\ref{eq:phase_shift})).

Within the stabilization method, we approximate the DOS using the discrete energy spectrum obtained in a finite box. For a fixed box size $L$, the density of states is formally given by
\begin{equation}
    \rho_L(E)=\sum_{N}\delta\big(E_N(L)-E\big)\,,
\end{equation}
where $E_N(L)$ are the discrete eigenenergies and $N$ labels the eigenstates. This expression reflects the fact that, in a finite volume, the spectrum is discrete rather than continuous. To obtain a smooth function of energy, one exploits the dependence of the eigenvalues on the box size $L$. When $L$ is chosen such that the boundary lies outside the interior region, the physical scattering properties are unaffected by further increases of $L$. In this regime, the DOS becomes effectively independent of $L$, and one can average over a small interval $\Delta L$ to smooth out the discrete spectrum,
\begin{equation}
    \langle \rho(E) \rangle=\frac{1}{\Delta L}\int_{L-\Delta L/2}^{L+\Delta L/2}\mathrm{d}L'\,\rho_{L'}(E)\,.
\end{equation}
This expression can be rewritten in terms of the $L$-dependence of the eigenvalues as
\begin{equation}
    \langle \rho(E) \rangle=\frac{1}{\Delta L}\sum_{N}
    \left|\frac{\mathrm{d}E_N(L')}{\mathrm{d}L'}\right|^{-1}_{E_N(L')=E}\,.
\end{equation}
This form provides a practical way to compute the DOS directly from the slopes of the stabilization curves $E_N(L)$. Regions where the energies vary slowly with $L$ (i.e., small derivatives) lead to large contributions to the DOS. This is precisely the signature of a resonance and explains why plateaus in the stabilization diagram translate into peaks in the DOS. Having obtained the averaged DOS, one can extract the resonance parameters by fitting it with a Lorentzian profile supplemented by a smooth background,
\begin{equation}\label{eq:lorentzian}
    f(E)=\frac{A_\mathrm{fit}}{(E-E_\mathrm{r})^2+\frac{\Gamma^2}{4}}+b_0+b_1 E\,.
\end{equation}
Here, $A_\mathrm{fit}$ is the amplitude of the resonant contribution, while $b_0$ and $b_1$ parametrize the non-resonant background.

The assumption of a background that changes as $b_0+b_1 E$ reflects the assumption that the underlying background typically changes only slowly with energy. Over the narrow fitting interval, this smooth contribution can therefore be well approximated by a linear function. This reduces the number of fit parameters while still capturing the essential behavior of the non-resonant contribution.

\subsection{Quasi-Bound Probability}

As a third approach, we introduce a method based directly on the spatial structure of the wave function. To the best of our knowledge, this method has not been discussed in the context of the stabilization method before. From a pedagogical perspective, this approach is appealing because it relies only on probability densities in position space, which students already encounter in introductory quantum mechanics. It therefore avoids the need to introduce more abstract constructs such as the density of states at an early stage.

The key physical idea of the method is that a resonance corresponds to a \emph{quasi-bound state}, i.e., a state in which the particle is temporarily localized in the interior region before eventually escaping. As a consequence, the probability density inside the interior region must be strongly enhanced when the energy matches the resonance energy.
{\color{black} To develop this idea, we derive an approximate relation for a quantity that we refer to, somewhat loosely, as the quasi-bound probability (QBP). The derivation assumes that, in the vicinity of an isolated resonance, a single interior state dominates and that the box size $L$ is sufficiently large (details are given in Appendix~\ref{app:B}). Under these conditions,}
\begin{equation}
\frac{\int_{\text{int}}\mathrm{d}x\,|\psi_{E}(x)|^2}{\int_{\text{ext}}\mathrm{d}x\,|\psi_{E}(x)|^2} \propto \frac{\cos^2(k a + \eta(E))}{(E-E_{\mathrm{r}})^2}\,,
\label{eq:quasi_bound_probability_first}
\end{equation}
where $\psi_{E}(x)$ denotes a finite-volume eigenfunction to Eq.~(\ref{eq:Schr_equation_1}) corresponding to energy $E$, and $\eta(E)$ is defined in Eq.~(\ref{eq:phase_shift}). As before, we divide the system somewhat arbitrarily into an \emph{interior region} (where the potential is non-zero) and an \emph{exterior region} (where the wave function behaves approximately like a free wave). 

 If the wave function is normalized to unity, the denominator can be eliminated, yielding the equivalent expression
\begin{equation}
    \left[\left(\int_{\text{int}}\mathrm{d}x\,|\psi_{E}(x)|^2\right)^{-1}-1\right]^{-1}\propto \frac{\cos^2(k a + \eta(E))}{(E-E_{\mathrm{r}})^2},
    \label{eq:quasi_bound_probability}
\end{equation}
which highlights that QBP is fully determined by the probability weight inside the interior region and exhibits a pronounced peak at the resonance energy. Consequently, resonance parameters can be extracted by fitting $\left[\left(\int_{\text{int}}\mathrm{d}x\,|\psi_{E}(x)|^2\right)^{-1}-1\right]^{-1}$ to the right-hand-side of Eq.~(\ref{eq:quasi_bound_probability}).

In contrast to phase‑shift- and DOS‑based approaches, the QBP method does not rely on derivatives of energy levels or on fitting nearly flat plateaus. Instead, it is based on an integrated quantity derived from the wave function itself, which can mitigate the impact of small numerical uncertainties. As demonstrated in the next section, the QBP extends the range of applicability of the stabilization method, at least in certain cases. A limitation of the QBP approach, however, is that it requires explicit access to the wave function, rather than relying solely on the energy spectrum -- a more transferable form of data that is generally better suited for numerical analysis~\cite{Langner2025}.

\section{Test Case: Delta Shell}
\label{sec:Delta_shell}
To benchmark the methods introduced above and to illustrate the stabilization method, we now work with a concrete and analytically tractable model system. Our goal is to provide clear examples that can be used in teaching. Furthermore, we want to demonstrate that the phase-shift, density-of-states, and quasi-bound probability approaches yield consistent resonance parameters.  {\color{black} Our examples also demonstrate that these methods are approximate and therefore do not necessarily reproduce the exact scattering parameters, even for well-isolated resonances.}

As a test case, we consider a one‑dimensional delta‑shell potential in the presence of an infinite wall:
\begin{align}
    V(x)=\begin{cases}
        \infty,&\,x\leq -a\,,\\
        U\delta(x),&x>-a\,.
    \end{cases}
\end{align}
The delta-shell model is particularly well suited for our purposes for several reasons. First, it supports resonances whose positions and widths can be obtained independently from the poles of the S-matrix that can be derived analytically (see Appendix~\ref{app:S_matrix}). This provides a reliable reference against which the results of the stabilization method can be compared. Second, the finite-volume problem admits analytic expressions for the eigenfunctions, while the eigenenergies can be computed efficiently with standard numerical methods. As a result, the model allows for a clear and controlled implementation of the stabilization procedure. Finally, in the context of decay processes, this model was studied by Winter~\cite{winter1961}, who demonstrated how it provides a clear time‑dependent interpretation of resonances, should such an analysis be required, {\color{black}see also Ref.~\cite{vachaspati2026} for the recent overview of the time-dependent problem.}

In the following, we first describe how stabilization diagrams are constructed for this system and how resonance parameters can be extracted from them. We then analyze both repulsive and attractive delta-shell potentials to systematically compare the three methods and to explore their behavior across different resonance regimes.

\begin{figure}[t]
    \centering
    \includegraphics[width=1\linewidth]{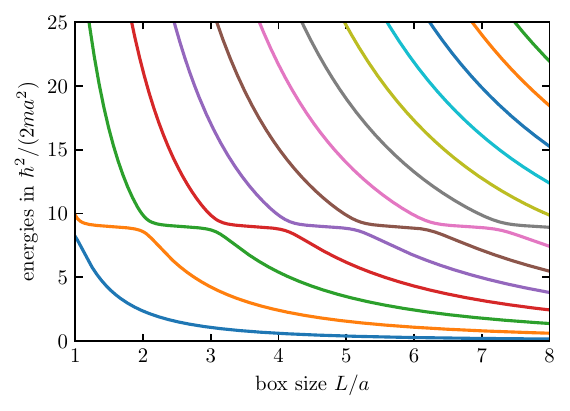}
    \caption{Stabilized energy levels of the repulsive delta-shell potential for $G=20$. The first resonance is clearly visible.}
    \label{fig:delta_shell_rep_levels}
\end{figure}

\begin{table}[t]
    \caption{Low-energy resonances of the repulsive delta-shell potential for different strengths $G$. Values are determined by finding S-matrix poles by numerically solving for the complex roots of Eq.~\eqref{eq:smat_denom}. Energies and widths are given in units of $\hbar^2/(2ma^2)$.}
    \label{tab:rep_vals_exact}

    \begin{ruledtabular}
    \begin{tabular}{c c c c c}
        $G$ & $E^{(1)}_\mathrm{r}$ & $\Gamma^{(1)}$ & $E^{(2)}_\mathrm{r}$ & $\Gamma^{(2)}$ \\
        \hline
        $20$ & $8.97$ & $0.246$ & $36.1$ & $1.79$ \\
        $10$ & $8.28$ & $0.766$ & $34.1$ & $4.82$ \\
        $5$  & $7.31$ & $1.93$  & $32.0$ & $10.0$ \\
    \end{tabular}
    \end{ruledtabular}
\end{table}

\subsection{Finite-Volume Delta Shell}
To apply the stabilization method, we introduce a second infinite wall at $x=L$, thereby confining the system to a finite interval $[-a,L]$. We refer to $L$ as the box size (note that $L$ is identical to the spatial extent of the exterior region). The corresponding finite-volume potential is thus given by
\begin{align}
    V_L(x)=\begin{cases}
        \infty,& x\leq -a\,,\\
        U\delta(x),& -a < x < L\,,\\
        \infty,& x\geq L\,.
    \end{cases}
\end{align}
We then solve the stationary Schr\"odinger equation for this confined system. For positive energies $E=\hbar^2 k^2/(2m)$, the solutions can be constructed piecewise. Imposing the boundary condition at $x=-a$ and matching across the delta potential at $x=0$, one obtains
\begin{equation}\label{eq:finite_eigenfunction}
\begin{aligned}
\psi_k&(x) = C\sin(kx+ka) \\
\quad &+ \begin{cases}
0, \quad -a < x \leq 0 \\
\frac{2mU}{k}\sin(ka)\sin(kx),\quad 0 < x \leq L\,,
\end{cases}
\end{aligned}
\end{equation}
where $C$ is a normalization constant.

To simplify the notation, we introduce dimensionless quantities for the coupling, momentum, and box size,
\begin{equation}\label{eq:dimless_units}
    G:=2mUa/\hbar^2\,, \qquad q:=ka\,, \qquad c:=\frac{L}{a}\,.
\end{equation}
The allowed momenta $q>0$ are then determined by the quantization condition arising from the boundary at $x=L$, which leads to
\begin{equation}
    q\sin(qc+q)+G\sin q\sin(qc)=0\,.
\end{equation}

This transcendental equation can be solved numerically. A convenient strategy is to start from the known solutions for $G=0$ (free particle in a box) and then continuously track the roots as $G$ is increased to the desired value. In this way, one obtains the discrete energy levels $E_N(L)$, where $N=1,2,3,\dots$ labels the eigenstates in ascending order of energy. The dependence of these energies on the box parameter $L$ defines the stabilization diagram.

\begin{figure*}[t]
    \centering
    
    \subfloat[]{
        \includegraphics[height=0.18\textheight]{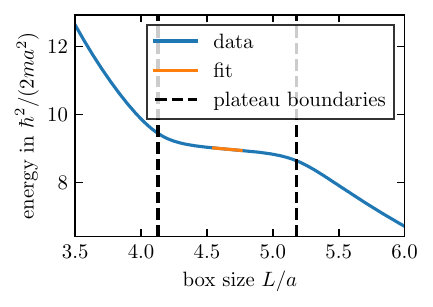}
    }\hspace{-0.01\linewidth}
    \subfloat[]{
        \includegraphics[height=0.18\textheight]{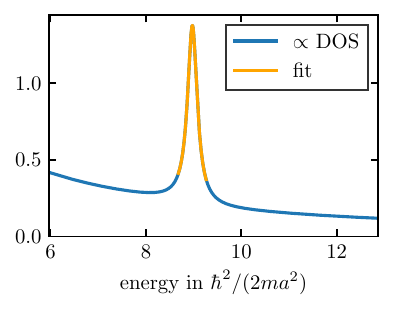}
    }\hspace{-0.01\linewidth}
    \subfloat[]{
        \includegraphics[height=0.18\textheight]{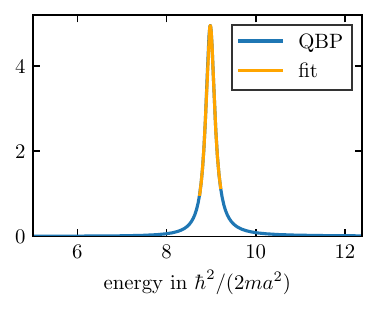}
    }
    
    \caption{Fit for all three methods for the first resonance of the repulsive delta-shell potential at $G=20$. (a) Fit to the $N=5$ energy level using $20\%$ of the plateau region (see supplementary material for details). (b) Fit to the density of states (DOS) obtained by averaging over the $N=8,9,10$ energy levels. (c) Fit to the quasi-bound probability ratio for the $N=10$ energy level.}
    \label{fig:fit_demos}
\end{figure*}

\begin{figure*}
    \centering
    
    \subfloat[]{
        \includegraphics[height=0.18\textheight]{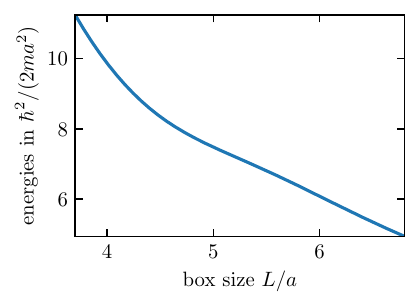}
    }\hspace{-0.01\linewidth}
    \subfloat[]{
        \includegraphics[height=0.18\textheight]{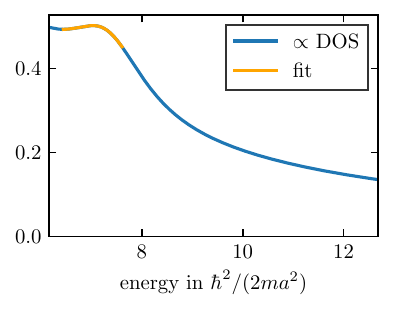}
    }\hspace{-0.01\linewidth}
    \subfloat[]{
        \includegraphics[height=0.18\textheight]{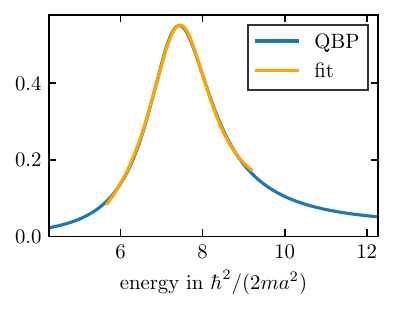}
    }
    
    \caption{Fit for all three methods for the first resonance of the repulsive delta-shell potential at $G=5$ (broad resonance). (a) The $N=5$ energy level; a plateau is not identifiable. (b) Fit to the density of states (DOS) obtained by averaging over the $N=8,9,10$ energy levels; the fit does not provide a reliable extraction. (c) Fit to the quasi-bound probability ratio for the $N=10$ energy level; this fit remains reliable.}
    \label{fig:fit_broad}
\end{figure*}

\subsection{Repulsive Delta Shell}\label{sec:rep_delta_shell}
In this section, we present results for a repulsive delta shell ($U>0$) for different coupling strengths. {\color{black} For each of the three methods, the state index $N$ should be chosen sufficiently large that the corresponding plateau occurs at a box size where finite-volume effects are small. For the DOS and QBP methods, shifting the analysis to larger $N$ has little effect on the extracted resonance parameters. By contrast, the direct fit to the energy levels shows a systematic dependence on $N$, with larger $N$ leading to increasing deviations, particularly in the extracted width. We therefore use the $N=5$ eigenstate for the direct fit to the energy levels. The fit is performed over the central 20\% of the plateau width. This choice represents a compromise between using a sufficiently large interval to constrain the fit reliably and remaining sufficiently close to the plateau center, where the approximation underlying the fitting function is most accurate.} For the quasi-bound probability (QBP) method, we use the $N=10$ eigenstate, and define the interior region as $x\in[-a,0]$. For the density-of-states (DOS) method, we average over the states $N=8,9,10$. Further details on the extraction procedures, their dependence on method-specific parameters, and the justification for choosing $N$ are provided in the supplementary material. These details can be relevant if the present paper is used as a basis for student's projects. 

For $U>0$, the potential is purely repulsive. It does not support true bound states but exhibits resonances, which appear as poles of the S-matrix in the complex plane (see Appendix~\ref{app:S_matrix}). While there are infinitely many such poles~\footnote{ The existence of infinitely many solutions follows from the periodic and meromorphic structure of $\cot q$ in Eq.~\eqref{eq:smat_denom}, which possesses an infinite sequence of poles and thus allows the pole condition to be satisfied repeatedly in their vicinity.}, their widths increase rapidly with energy, such that higher-lying poles lose their clear resonance character and become difficult to resolve in practice.

The stabilization pattern for $G=20$ is shown in Fig.~\ref{fig:delta_shell_rep_levels}, where the lowest-lying resonance is clearly visible as a pronounced plateau. We focus on the first two resonances for coupling strengths $G=5,10,20$. The corresponding resonance energies and widths, obtained from the S-matrix, are listed in Table~\ref{tab:rep_vals_exact}. As the coupling $G$ decreases, the resonances become broader, providing a useful test of the robustness of the different extraction methods.

\begin{table*}
    \caption{First and second resonances of the repulsive delta-shell potential for different strengths $G$ extracted via the stabilization method. Values from all three methods are presented. Numbers are omitted where the respective method was not able to extract them reliably. Energies and widths are given in units of $\hbar^2/(2ma^2)$.}
    \label{tab:rep_vals_methods}

    \begin{ruledtabular}
    \begin{tabular}{c c c c c c c c c c c c c}
        $G$ 
        & \multicolumn{6}{c}{First resonance} 
        & \multicolumn{6}{c}{Second resonance} \\
        
        & $E_{\mathrm{r,fit}}$ & $\Gamma_\mathrm{fit}$
        & $E_{\mathrm{r,dos}}$ & $\Gamma_\mathrm{dos}$
        & $E_{\mathrm{r,qbp}}$ & $\Gamma_\mathrm{qbp}$
        & $E_{\mathrm{r,fit}}$ & $\Gamma_\mathrm{fit}$
        & $E_{\mathrm{r,dos}}$ & $\Gamma_\mathrm{dos}$
        & $E_{\mathrm{r,qbp}}$ & $\Gamma_\mathrm{qbp}$ \\
        \hline
        
        $20$ & $8.97$ & $0.258$ & $8.97$ & $0.246$ & $8.98$ & $0.245$
             & $36.1$ & $1.91$  & $36.1$ & $1.78$  & $36.2$ & $1.77$ \\
        
        $10$ & $8.26$ & $0.892$ & $8.27$ & $0.746$ & $8.30$ & $0.759$
             & $34.2$ & $4.92$  & $34.0$ & $4.64$  & $34.3$ & $4.75$ \\
        
        $5$  & / & / & / & / & $7.43$ & $1.90$
             & / & / & / & / & $32.6$ & $9.65$ \\
        
    \end{tabular}
    \end{ruledtabular}
\end{table*}

The results obtained with the three methods are summarized in Table~\ref{tab:rep_vals_methods}. As an illustrative example, Fig.~\ref{fig:fit_demos} shows fits for all three methods in the case $G=20$. For $G=20$ and $G=10$, all methods yield consistent results: the extracted resonance energies agree with the exact values to within less than one percent. The widths exhibit larger deviations, with the phase-shift-fit method showing the largest errors. {\color{black} These errors are consistent with our quantitative assessment of the uncertainties associated with the fitting procedure, see the next subsection. We conclude that, whereas the resonance energy is determined primarily by the position of the plateau, the width depends more sensitively on its detailed shape and is therefore more susceptible to the higher-order corrections neglected in the second line of Eq.~(\ref{eq:phase_shift_fit_linearization}). The dependence on the chosen fitting interval can also affect the accuracy of the results as discussed in the Supplementary Material, leading to the largest systematic uncertainties of the phase-shift-fit method. }

For the broader resonances at $G=5$, clear differences between the methods emerge (see Fig.~\ref{fig:fit_broad}). The phase-shift-fit method fails due to the absence of a well-defined plateau, making a reliable identification of the fitting region impossible. Similarly, in the DOS approach, the resonance peak becomes too weak to be clearly distinguished from the background. In contrast, the QBP method continues to produce a pronounced signal, allowing for a meaningful extraction of resonance parameters.

One can therefore conclude that, while all three methods perform well for narrow resonances, the QBP method is more robust in the regime of broad resonances, where the signatures in the energy spectrum and DOS become less distinct. {\color{black} Formal fit uncertainties for both the repulsive and attractive cases are
discussed together in Sec.~\ref{sec:attr_delta_shell}.}

\subsection{Attractive Delta Shell}\label{sec:attr_delta_shell}
A repulsive delta shell naturally generates a high barrier that partitions space into interior and exterior regions. By contrast, for an attractive delta shell such a separation is not immediately apparent, making it instructive to examine this case as well. We find that the attractive delta shell ($U<0$) displays resonant features that are qualitatively similar to those of the repulsive case  (see Appendix~\ref{app:attractive_delta}), albeit with important additional effects associated with bound states~\footnote{
In the infinite-volume limit ($L\to\infty$), the potential supports a single bound state for $G<-1$, while for $G=-1$ a zero-energy state occurs. In the range $-1<G<0$, no bound state exists. In this regime, the system exhibits a \emph{virtual state} instead -- a pole of the S-matrix on the negative imaginary axis of the momentum plane (recall that a bound state appears on the positive imaginary axis). 
Unlike a bound state, a virtual state is not associated with a normalizable wave function, 
but it can still strongly affect low-energy scattering.}. The underlying physical reason for this similarity in the strong-coupling regime is the presence of a well-localized ground state, which enforces the vanishing of the wave function for the low-lying excited states at the location of the potential in order to maintain orthonormality of the spectrum.

{\color{black} Combining the repulsive results discussed above with the corresponding attractive results presented in Appendix~\ref{app:attractive_delta}, we also assess the fit stability quantitatively. We estimated formal $1\sigma$ uncertainties from the covariance matrices of the nonlinear least-squares fits. For the reliably extracted resonances at $G=\pm10,\pm20$, the relative uncertainties in $E_{\mathrm{r}}$ remain below $0.08\%$, $0.004\%$, and $0.007\%$ for the phase shift, DOS, and QBP methods, respectively. The corresponding uncertainties in $\Gamma$ remain below $1.8\%$, $0.16\%$, and $0.24\%$. Thus, the formal fit uncertainties are largest for the phase shift, particularly for the resonance width, while the DOS and QBP fits show substantially smaller uncertainties. For the broad resonances at $G=\pm5$, where only the QBP method yields a reliable extraction, its relative fit uncertainties remain below $0.02\%$ in $E_{\mathrm{r}}$ and $0.4\%$ in $\Gamma$. We emphasize that these covariance-based uncertainties characterize the local stability of a given fit and do not account for systematic effects or for a failure of the assumed fit model.}

\section{Conclusions}
In this work, we have discussed the stabilization method as a framework for introducing scattering resonances. By sidestepping an explicit treatment of the continuum and instead focusing on finite-volume quantities, the method offers an intuitive route to understanding resonant phenomena. This perspective may help bridge the conceptual gap that is often encountered when resonances are first introduced in quantum mechanics.

We showed how the stabilization method naturally encodes scattering information through the dependence of discrete energy levels on the box size. By linking this behavior to the phase shift, the density of states, and the spatial localization of the wave function, the three methods offer different but closely related perspectives on resonances. In particular, all approaches ultimately rely on the characteristic enhancement associated with quasi-bound states, either in energy space (plateaus, density of states) or directly in position space (probability densities).

Using the one-dimensional delta-shell potential as an accessible toy model, we illustrated the methods. For narrow resonances, all three approaches yield consistent resonance energies and reasonably accurate widths. However, clear differences emerge for broader resonances: both the phase-shift-fit and DOS methods become increasingly unreliable due to the lack of well-defined plateaus and weakly pronounced DOS peaks. In contrast, the QBP method remains applicable and provides a robust signal even in this regime.

This paper can serve as a stepping stone toward a range of research-intensive undergraduate projects. Examples include: (i) the calculation of resonances in experimentally relevant one-dimensional systems~\cite{Mistakidis2023}, with related theoretical discussions available in Refs.~\cite{Hunn2013,Lundmark2015,Gharashi2015}; and (ii) the calculation of scattering properties in harmonic traps~\cite{Fedorov2009,Fedorov2025}, which are often more amenable to numerical implementation. {\color{black} For central interactions, where the partial waves decouple, both the stabilization framework and the QBP diagnostic generalize straightforwardly to radial three-dimensional problems. In fact, the considered delta-shell problems can be directly used to study the $s$-wave three-dimensional problem. }

Finally, we note that from a practical perspective, our findings suggest that the QBP method allows one to extract properties of the resonance where other methods fail. A more detailed analysis of the QBP method, including its theoretical foundations and limitations, would be a valuable direction for future research. {\color{black} In particular, a natural next step is to extend the QBP method to multichannel systems with coupled channels.}

\begin{acknowledgments}
We thank Dmitri Fedorov for discussions and Hans Fynbo, Nathan Harshman, Timothy Backert and Philipp Quoß for comments on the manuscript. 
D.K. and H.W.H. acknowledge support by the Deutsche Forschungsgemeinschaft (DFG, German Research Foundation) - Project-ID 279384907 - SFB 1245.
H.W.H. was supported by the BMFTR Contract 05P24RDB. A.G.V. was supported in part by a visiting professorship from the ExtreMe Matter Institute EMMI at the GSI Helmholtzzentrum für Schwerionenforschung, Darmstadt, Germany.
A.G.V. was also supported in part by
the Novo Nordisk Foundation (grant reference number NNF25OC0102659).
\end{acknowledgments}

\appendix   

\section{Derivation of Eq.~(\ref{eq:coupling_strong})}
\label{app:strong}
Here, we discuss the derivation of Eq.~(\ref{eq:coupling_strong}). Our discussion is based upon the results known in the literature~\cite{Volosniev2013,Gharashi2015}. 
Using the Hellmann-Feynman theorem we can derive the derivative of the energy with respect to the interaction strength, $U$:
\begin{equation}
\frac{\partial E}{\partial U}=\frac{\langle \psi|\delta(x)|\psi\rangle}{\langle \psi|\psi\rangle}\,.
\end{equation}
The numerator in this expression can be calculated using the delta-function boundary condition
\begin{equation}
U\psi(0)=\frac{\hbar^2}{2m}\left(\frac{\partial \psi}{\partial x}{\Big|}_{x=a+}-\frac{\partial \psi}{\partial x}{\Big|}_{x=a-}\right)\,,
\end{equation}
where $a+$ ($a-$) implies the right (left) derivative.  

From the derivative of the energy, we compute 
\begin{equation}
E\simeq \frac{\langle \psi|H_{\frac{1}{U}=0}|\psi\rangle}{\langle \psi|\psi\rangle}-\frac{\hbar^4}{4 m^2 U}\frac{\left(\frac{\partial \psi}{\partial x}|_{x=a+}-\frac{\partial \psi}{\partial x}|_{x=a-}\right)}{\langle\psi|\psi\rangle}\,.
\end{equation}
The function $\psi$ is taken in the limit $1/U=0$ where we can write $\psi=a_1\psi^{(1)}_{\mathrm{int}}+a_2 \psi^{(1)}_{\mathrm{ext}}+a_3\psi^{(2)}_{\mathrm{ext}}$. The coefficients $a_i$ are found by minimizing the energy, i.e., $\partial E/\partial a_i=0$. This procedure allows us to estimate the couplings, see Eq.~(\ref{eq:coupling_strong}). Note that the minimization procedure leads to an effective Hamiltonian that is somewhat different from   Eq.~(\ref{eq:hamiltonian_toy}). In particular, there is a $1/U$-correction to the diagonal elements. We disregard this in the qualitative discussion of Sec.~\ref{stab_method_intro} since it does not affect our main conclusions. 

\section{Derivation of Eq.~(\ref{eq:quasi_bound_probability_first})}
\label{app:B}
The solution to Eq.~(\ref{eq:Schr_equation_1}) in the interior region can be found using the expansion (see Sec. IV of Ref.~\cite{Lane1958}) 
\begin{equation}
\psi_E(x)=\sum_\lambda \frac{u_\lambda(x)u_\lambda(a)}{E_\lambda-E}\left(\frac{\mathrm{d}\psi_E}{\mathrm{d}x}\right)_{x=a},
\label{eq:expansion}
\end{equation}
where $\{u_{\lambda}\}$ is an orthonormal set of solutions to Eq.~(\ref{eq:Schr_equation_1}) in the interior region subject to the condition $\mathrm{d}u_\lambda/\mathrm{d}x|_{x=a}=0$. Equation~(\ref{eq:expansion}) allows us to compute the probability 
\begin{equation}
\int_\mathrm{int}\mathrm{d}x |\psi_E(x)|^2=\left(\frac{\mathrm{d}\psi_E}{\mathrm{d}x}\right)^2_{x=a}\sum_\lambda \frac{u^2_\lambda(a)}{(E_\lambda-E)^2}.
\label{eq:intergral_int}
\end{equation}
In the vicinity of the resonance, a single term dominates the sum in Eq.~(\ref{eq:intergral_int}):
\begin{equation}
\int_\mathrm{int}\mathrm{d}x |\psi_E(x)|^2\propto \left(\frac{\mathrm{d}\psi_E}{\mathrm{d}x}\right)^2_{x=a}\frac{1}{(E_\mathrm{r}-E)^2},
\end{equation}
where the derivative can be calculated using Eq.~(\ref{eq:boundary_condition}): 
\begin{equation}
\left(\frac{\mathrm{d}\psi_E}{\mathrm{d}x}\right)^2_{x=a}=A^2 k^2\cos^2(ka+\eta(E)).
\end{equation}
Here, $A^2$ is proportional to $\int_{\mathrm{ext}}\mathrm{d}x|\psi_E(x)|^2$ in the limit of large values of $L$. Indeed, 
\begin{equation}
A^2=\frac{\int_{\mathrm{ext}}\mathrm{d}x|\psi_E(x)|^2}{\int_{\mathrm{ext}}\mathrm{d}x \sin^2(kx+\eta(E))}\simeq \frac{2}{L}\int_{\mathrm{ext}}\mathrm{d}x|\psi_E(x)|^2.
\end{equation}
By combining the equations above, we derive Eq.~(\ref{eq:quasi_bound_probability_first}). 

\section{S-matrix}
\label{app:S_matrix}

To benchmark the extraction of resonance parameters, we define the resonances via poles of the S-matrix. For the present one-dimensional system, all incoming flux is reflected at the infinite wall, such that the S-matrix reduces to the reflection amplitude.
{\color{black} The latter can be obtained directly from the finite-volume solution given in Eq.~\eqref{eq:finite_eigenfunction}. Before imposing the boundary condition at $x=L$, the exterior part of this wave function already satisfies the Schrödinger equation and the matching condition at the delta shell.
Removing the wall at $x=L$ therefore simply allows this solution to be continued to infinity. Using the dimensionless quantities introduced in Eq.~\eqref{eq:dimless_units}, the wave function for $x>0$ can be written as
\begin{equation}
    \psi_k(x)
    = C\left[
        \sin(kx+q)
        + \frac{G}{q}\sin q\,\sin(kx)
      \right].
\end{equation}
Introducing
\begin{equation}
    X(q)=q\cot q+G,
\end{equation}
and rewriting the trigonometric functions in terms of incoming and outgoing plane waves gives
\begin{equation}
    \psi_k(x)=
    \frac{C\sin q}{2q}
    \left[
        \bigl(q+\mathrm{i}X(q)\bigr)\e^{-\mathrm{i}kx}
        +
        \bigl(q-\mathrm{i}X(q)\bigr)\e^{\mathrm{i}kx}
    \right].
\end{equation}
The S-matrix is the ratio of the outgoing to the incoming amplitude,
and hence
\begin{equation}\label{eq:smat}
    S(q)=
    \frac{q-\mathrm{i}X(q)}
         {q+\mathrm{i}X(q)}
    =
    \frac{\mathrm{i}q+q\cot q+G}
         {\mathrm{i}q-q\cot q-G}\,.
\end{equation}}
Resonances correspond to poles of $S(q)$ in the complex momentum plane, which are given by the zeros of the denominator,
\begin{equation}\label{eq:smat_denom}
    \mathrm{i} q-q\cot q-G=0\,.
\end{equation}

\begin{figure}[t]
    \centering
    \includegraphics[width=1\linewidth]{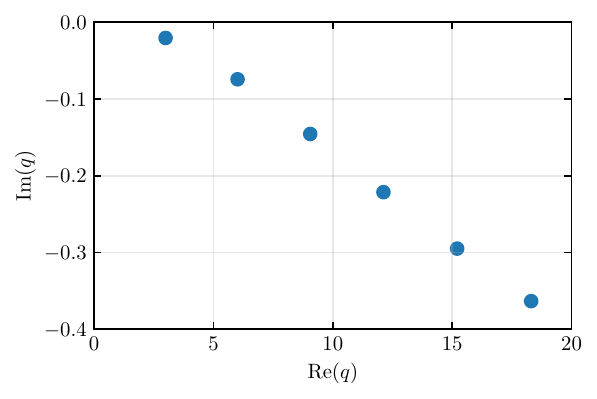}
    \caption{The first six lowest lying poles of the S-matrix in the complex momentum plane for $G=20$.}
    \label{fig:poles}
\end{figure}

These complex roots can be determined numerically, for instance by using standard root-finding algorithms in the complex plane.  A pole at complex momentum $q_0$ corresponds to a resonance with complex energy $E_0 = \hbar^2q_0^2/(2ma^2)$. Writing $E_0 = E_\mathrm{r} - \mathrm{i}\Gamma/2$ defines the resonance energy $E_\mathrm{r}$ and width $\Gamma$. A plot of the lowest lying poles is given in Fig.~\ref{fig:poles}.

\begin{figure}[tb]
    \centering
    \includegraphics[width=1\linewidth]{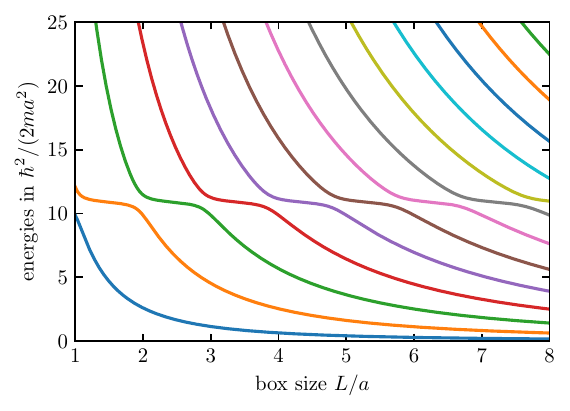}
    \caption{Stabilized energy levels of the attractive delta-shell potential for $G=-20$. The first resonance is clearly visible.}
    \label{fig:delta_shell_attr_levels}
\end{figure}

\begin{table}[tb]
    \caption{First and second resonances of the attractive delta-shell potential for different strengths $G$. Values are determined by numerically solving for the complex roots of Eq.~\eqref{eq:smat_denom}. Energies and widths are given in units of $\hbar^2/(2ma^2)$.}
    \label{tab:attr_vals_exact}

    \begin{ruledtabular}
    \begin{tabular}{c c c c c}
        $G$ & $E^{(1)}_\mathrm{r}$ & $\Gamma^{(1)}$ & $E^{(2)}_\mathrm{r}$ & $\Gamma^{(2)}$ \\
        \hline
        $-20$ & $10.9$ & $0.357$ & $43.2$ & $2.44$ \\
        $-10$ & $11.8$ & $1.43$  & $45.3$ & $7.23$ \\
        $-5$  & $12.8$ & $4.32$  & $46.7$ & $15.1$ \\
    \end{tabular}
    \end{ruledtabular}
\end{table}

\begin{widetext}
\begin{table*}[tb]
    \caption{First and second resonances of the attractive delta-shell potential for different strengths $G$ extracted via the stabilization method. Values from all three methods are presented. Energies and widths are given in units of $\hbar^2/(2ma^2)$.}
    \label{tab:attr_vals_methods}

    \begin{ruledtabular}
    \begin{tabular}{c c c c c c c c c c c c c}
        $G$ 
        & \multicolumn{6}{c}{First resonance} 
        & \multicolumn{6}{c}{Second resonance} \\
        
        & $E_{\mathrm{r,fit}}$ & $\Gamma_\mathrm{fit}$
        & $E_{\mathrm{r,dos}}$ & $\Gamma_\mathrm{dos}$
        & $E_{\mathrm{r,qbp}}$ & $\Gamma_\mathrm{qbp}$
        & $E_{\mathrm{r,fit}}$ & $\Gamma_\mathrm{fit}$
        & $E_{\mathrm{r,dos}}$ & $\Gamma_\mathrm{dos}$
        & $E_{\mathrm{r,qbp}}$ & $\Gamma_\mathrm{qbp}$ \\
        \hline
        
        $-20$ & $10.9$ & $0.378$ & $10.9$ & $0.357$ & $10.9$ & $0.360$
              & $43.2$ & $2.64$  & $43.2$ & $2.43$  & $43.2$ & $2.47$ \\
        
        $-10$ & $11.8$ & $1.56$  & $11.8$ & $1.38$  & $11.9$ & $1.44$
              & $45.6$ & $7.24$  & $45.2$ & $6.92$  & $45.6$ & $7.32$ \\
        
        $-5$  & \textemdash & \textemdash & \textemdash & \textemdash & $13.1$ & $4.34$
              & \textemdash & \textemdash & \textemdash & \textemdash & $47.7$ & $14.9$ \\
        
    \end{tabular}
    \end{ruledtabular}
\end{table*}
\end{widetext}

\section{Attractive delta-shell potential}
\label{app:attractive_delta}

The stabilization pattern for $G=-20$ is shown in Fig.~\ref{fig:delta_shell_attr_levels}, where the lowest-lying resonance is visible as a plateau. As in the repulsive case, we analyze the first two resonances for coupling strengths $G=-5,-10,-20$. The corresponding reference values for the resonance energies and widths, obtained from the S-matrix, are listed in Table~\ref{tab:attr_vals_exact}. As for the repulsive case, decreasing $|G|$ leads to broader resonances.

The results obtained with the three extraction methods are summarized in Table~\ref{tab:attr_vals_methods}. For $G=-20$ and $G=-10$, all methods yield consistent resonance energies with deviations below one percent, while the extracted widths show larger uncertainties. For the broader resonances at $G=-5$, the same limitations as in the repulsive case arise: the phase-shift-fit method fails due to the absence of a clearly identifiable plateau, and the DOS method suffers from a poorly resolved peak. In contrast, the QBP method again provides a comparatively clear signal, enabling a meaningful extraction of resonance parameters.

Overall, the attractive delta shell confirms the conclusions drawn from the repulsive case: while all methods perform well for narrow resonances, the QBP method remains the most robust approach when dealing with broad resonances.

\newpage

\bibliography{bibliography.bib}

\end{document}